\documentclass[12pt,a4paper]{article}
\usepackage{epsfig}
\usepackage{mathtools}
\usepackage{amsfonts}
\usepackage{amssymb}
\usepackage{amsmath}
\usepackage{color}
\usepackage{slashed}
\usepackage{cite}
\usepackage{caption}
\usepackage{subcaption}

\usepackage{geometry} 
\usepackage{graphicx}
\graphicspath{{pictures/}}

\usepackage{multirow}           
\usepackage{array}              

\def\Li2  {\text{Li}_2}
\def\LL   {\mathcal L}
\def\ep {\varepsilon}
\def\im {\text{Im}}
\def\tr {\text{tr}}
\def\bk {\mathbf{k}}
\def\erf {\mbox{erf}}
\def\unren {\text{unrenom}} 
\def\MSbar {\overline{MS}} 
\def\gE {\gamma_E}

\begin{document}

\begin{center}

\vspace{3cm}

{\bf \Large Two-loop corrections to false vacuum decay \\ in scalar field theory} \vspace{1cm}

{\large M.A. Bezuglov$^{1,2}$ and A.I. Onishchenko$^{1,2,3}$}\vspace{0.5cm}

{\it $^1$Bogoliubov Laboratory of Theoretical Physics, Joint
	Institute for Nuclear Research, Dubna, Russia, \\
	$^2$Moscow Institute of Physics and Technology (State University), Dolgoprudny, Russia, \\
	$^3$Skobeltsyn Institute of Nuclear Physics, Moscow State University, Moscow, Russia}\vspace{1cm}

\abstract{We consider radiative corrections to false vacuum decay in a four-dimensional scalar field theory with cubic and quartic potential. Using planar thin wall approximation we were able to get analytical expression for the decay rate up to two loop order. The results obtained employ dimensional regularization and $\MSbar$ renormalization scheme.}
\end{center}

\begin{center}
\end{center}

\newpage

\section{Introduction}

Recently, first-order phase transitions driven by scalar fields attracted a lot of attention in high-energy, astro-particle physics and cosmology communities.  In particular, a lot of recent research was connected with the possible metastability\footnote{For tunneling rates calculations in Standard Model and its extensions see \cite{SMtunneling1,SMtunneling2,SMtunneling3,SMtunneling4,SMtunneling5,SMtunneling6,SMtunneling7,SMtunneling8,SMtunneling9,
SMtunneling10,SMtunneling11,SMtunneling12,SMtunneling13,SMtunneling14}
 and references therein.} of Standard Model vacuum at a scale around $10^{11}$ GeV \cite{SMmetastability1,SMmetastability2,SMmetastability3,SMmetastability4,SMmetastability5,SMmetastability6,SMmetastability7,SMmetastability8}.
	
The systematic description of this type of quantum transitions in the framework of quantum field theory first appeared in \cite{Coleman1,Coleman2,Coleman3,Kobzarev}. At present we already have a number of different approaches, such as potential deformation method \cite{KleinertPathIntegrals,IntroQuantumMechanics,QFTCriticalPhenomena,InstantonsLargeN,WeinbergClassicalSolutions,PrecisionDecayRates} and direct computation via path integrals \cite{DirectApproachQuantumTunneling,PrecisionDecayRates}. The one-loop computations are the most developed. To evaluate functional determinants we may choose from direct evaluation of the spectrum for solvable potentials\footnote{See \cite{InstantonsLargeN} and references therein.}, heat kernel methods \cite{heatkernel-usermanual,CalculationsExternalFields,MassiveContributionsQCDtunneling,DunneFunctionalDeterminants}, Green function methods \cite{KleinertChervyakov1,KleinertChervyakov2,GreenFunctionScalarQFT,GarbrechtMillington2,GarbrechtMillington3,GarbrechtMillington4} or Gel'fand-Yaglom method \cite{GenfaldYaglom} and its generalizations \cite{KirstenMcKane1,KirstenMcKane2}. Beyond one loop the corresponding techniques were mostly developed in the study of instantons \cite{ABCinstantons,InstantonsQCD,LecturesInstantons} in quantum mechanics \cite{Instantons2loop,Olejnik,Instantons3loop,Instantons3loop-SineGordon,QuantumThermalFluctuationsQM,VacuumDecayQM}, investigation of effective Euler-Heisenberg lagrangian \cite{EulerHeisenberg1,EulerHeisenberg2,EulerHeisenberg3} and computation of quantum corrections to classical string solutions \cite{Tseytlin1,Tseytlin2}.  

The purpose of this paper is to extend the methods of \cite{Instantons2loop,Olejnik,Instantons3loop,Instantons3loop-SineGordon,QuantumThermalFluctuationsQM,VacuumDecayQM} and \cite{KleinertChervyakov1,KleinertChervyakov2,GreenFunctionScalarQFT,GarbrechtMillington2,GarbrechtMillington3,GarbrechtMillington4} for the computation of higher order radiative corrections to false vacuum decay in four-dimensional scalar field theory. Here we will concentrate on two-loop corrections within dimensional regularization and $\MSbar$ scheme. The generalization to Coleman-Weinberg (CW) scheme and cut-off regularization will be presented in our subsequent paper. The paper is organized as follows. The section \ref{VacuumDecay} and its subsections contain all the material of the present paper. First, following \cite{GreenFunctionScalarQFT} we describe our model with single scalar field experiencing cubic and quartic self-interactions and false vacuum decay in it at one-loop order. Next, in subsection \ref{GreenFunctionsubsection} we present required Green functions in the bounce background and false vacuum in a planar thin wall approximation. Most of these results are already known from \cite{GreenFunctionScalarQFT}, expect the expression of bounce Green function in a special kinematical limit which is required in a subsequent Feynman diagrams calculation. Then, in subsections \ref{OneLoopDecay} and \ref{TwoLoopDecay}  we present the details of one and two loop calculations in dimensional regularization. Finally in section \ref{Conclusion} we come with our conclusion. The \ref{sunset-appendix} contains the details of calculation of the most complicated two-loop sunset diagram.

\section{False vacuum decay in scalar field theory}\label{VacuumDecay}

Let us consider false vacuum decay in a four-dimensional field theory with a single real scalar field $\Phi\equiv \Phi (x)$ and a lagrangian\footnote{We refer the reader to \cite{GreenFunctionScalarQFT} for more details.} 
\begin{equation}
\LL = \frac{1}{2}(\partial_{\mu}\Phi)^2 + U\, ,
\end{equation}
where 
\begin{equation}
U = -\frac{1}{2}\mu^2 \Phi^2 + \frac{g}{3!}\Phi^3 + \frac{\lambda}{4!}\Phi^4 + U_0
\end{equation}
The potential $U$ has two minima $\varphi = v_{\pm}$ with separation $\triangle v = v_{+} - v_{-}$ and difference in potential levels $\triangle U = U_{v_{+}} - U_{v_{-}} = 2\epsilon$. When $g\to 0$ the minima at $v_{\pm} = \pm v$ become degenerate as $\epsilon\sim \frac{g v^2}{6}\to 0$. It is precisely this limit which allows analytical treatment and which was considered in \cite{GreenFunctionScalarQFT}. In what follows it is convenient to chose $U_0 = \frac{\mu^2 v^2}{4} - \frac{g v^3}{6}$, so that the potential vanishes in the false vacuum $\varphi = +v$.

The false vacuum decay in this theory accounting for semi-classical tunneling between false ($\varphi = v$) and true ($\varphi = -v$) vacua together with first quantum corrections was first considered in \cite{Coleman1,Coleman2}. In a path integral formulation the false vacuum decay probability is given by the ratio of the path integrals evaluated around bounce and false vacuum solutions. In the present model the bounce corresponds to a four dimensional bubble of some radius $R$ separating true vacuum inside bubble from a false vacuum outside. The latter is given by a $O(4)$ symmetric solution  of the classical equations of motion
\begin{equation}
-\partial^2\varphi + U' (\varphi) = 0\, , \label{eqmotion}
\end{equation}
where $'$ denotes the derivative with respect to the field $\varphi$. Rewriting the above equation in hyperspherical coordinates
\begin{equation}
-\frac{d^2}{dr^2}\varphi - \frac{3}{r}\frac{d}{dr}\varphi + U' (\varphi) \label{eqmotion-hyperspherical}
\end{equation}
and using thin-wall approximation, that is neglecting both cubic self-interaction $g\phi^3$ and damping term in (\ref{eqmotion-hyperspherical}), it is easy to see that the bounce is given by the well-known kink solution \cite{NonperturbativeMethodsReview}:
\begin{equation}
\varphi (r) = v \tanh [\gamma (r-R)]\, , \quad \gamma = \frac{\mu}{\sqrt{2}}\, , \quad v= 2\gamma\sqrt{\frac{3}{\lambda}}
\end{equation}
The radius of the bubble is then obtained by extremizing the bounce action \cite{GreenFunctionScalarQFT}:
\begin{equation}
R = \frac{12\gamma}{g v} = \frac{2\sqrt{3\lambda}}{g}\, .
\end{equation}
The classical action itself evaluated at bounce solution is given by
\begin{equation}
S_{b} = \int d^4 x \left[
\frac{1}{2}\left(\frac{d\varphi}{d r}\right)^2 + U(\varphi) 
\right] = \frac{8\pi^2 R^3\gamma^3}{\lambda}\, .
\end{equation}
Next, introducing partition function
\begin{equation}
Z [J] = \int [d\Phi] \exp \left[
-\frac{1}{\hbar} \left(
S [\Phi] - \int d^4 x J(x) \Phi (x)
\right)
\right]
\end{equation}
the false vacuum decay rate could be written as
\begin{equation}
\Gamma = -2 \im E_{FV}/\hbar = 2 | \im Z[0]|/T\, ,
\end{equation}
where $T$ is the Euclidean time of the bounce and partition function being evaluated around bounce solution. Note, that here the partition function should be evaluated in one bounce approximation, see \cite{Coleman2,SMtunneling3} for more details. The latter goes through saddle point approximation  and at one-loop order is given by \cite{GreenFunctionScalarQFT}:
\begin{equation}
i Z[0] = e^{-S_{b}/\hbar} \left|
\frac{\lambda_0 \det^{(5)} G^{-1} (\varphi)}{\frac{1}{4}(V T)^2 \left(\frac{S_{b}}{2\pi\hbar}\right)^4 (4\gamma^2)^5 \det^{(5)} G^{-1} (v)}
\right|^{-1/2}\, , \label{Z1loop}
\end{equation}
where $\lambda_0 = -\frac{3}{R^2}$ is the negative eigenvalue of $G^{-1} (\varphi)$ operator and  $\det^{(5)}$ denotes the determinant calculated only over the continuum of positive-definite eigenvalues, that is omitting zero and negative eigenmodes. Here and below, the parameter $\hbar$ is introduced simply in order to count loops, the actual perturbation theory is constructed as series in $\lambda$.  

The inverse of Green functions $G^{-1}$ at the bounce $\varphi$ and in false vacuum $v$ are defined as  ($\triangle^{(4)}$ is the four-dimensional Laplacian):
\begin{equation}
G^{-1}(\varphi)\equiv \frac{\delta^2 S[\Phi]}{\delta\Phi^2 (x)}\Bigg|_{\Phi = \varphi} = -\triangle^{(4)} + U'' (\varphi)\, ,
\end{equation}
The spectrum of operator $G^{-1}(\varphi)$ at bounce solution
\begin{equation}
(-\triangle^{(4)} + U'' (\varphi))\phi_{n j} = \lambda_{n j} \phi_{n j}
\end{equation}
is given by \cite{GreenFunctionScalarQFT}:
\begin{equation}
\lambda_{nj} = \gamma^2 (4-n^2) + \frac{j(j+2)-3}{R^2}
\end{equation}
and contains one negative mode at $\lambda_0 = \lambda_{20}$ and four zero modes at $\lambda_{21}$. The "continuum" of positive-definite modes starts at $\lambda_{10}\approx \lambda_{11} = 2\gamma^2$. 

Knowing the expression for the partition function evaluated at bounce solution \eqref{Z1loop} the expression for false vacuum decay (tunneling probability per unit volume) at one-loop order takes the form \cite{GreenFunctionScalarQFT}:
\begin{equation}
\frac{\Gamma}{V} = \left(\frac{S_b}{2\pi\hbar}\right)^2 \frac{(2\gamma)^5 R}{\sqrt{3}} \exp \left[
-\frac{1}{\hbar} S_b + I^{(1)}
\right]\, , \label{decay1loop}
\end{equation}
where
\begin{equation}
I^{(1)} = -\frac{1}{2}\tr^{(5)} \left(\ln G^{-1}(\varphi) - \ln G^{-1} (v)  \right) \label{funcdet1loop}
\end{equation}
and $\tr^{(5)}$ denotes the trace only over positive-definite eigenmodes.

\subsection{Green function in bounce background}\label{GreenFunctionsubsection}

\begin{figure}[h]
	\center{\includegraphics[width=0.35\textwidth]{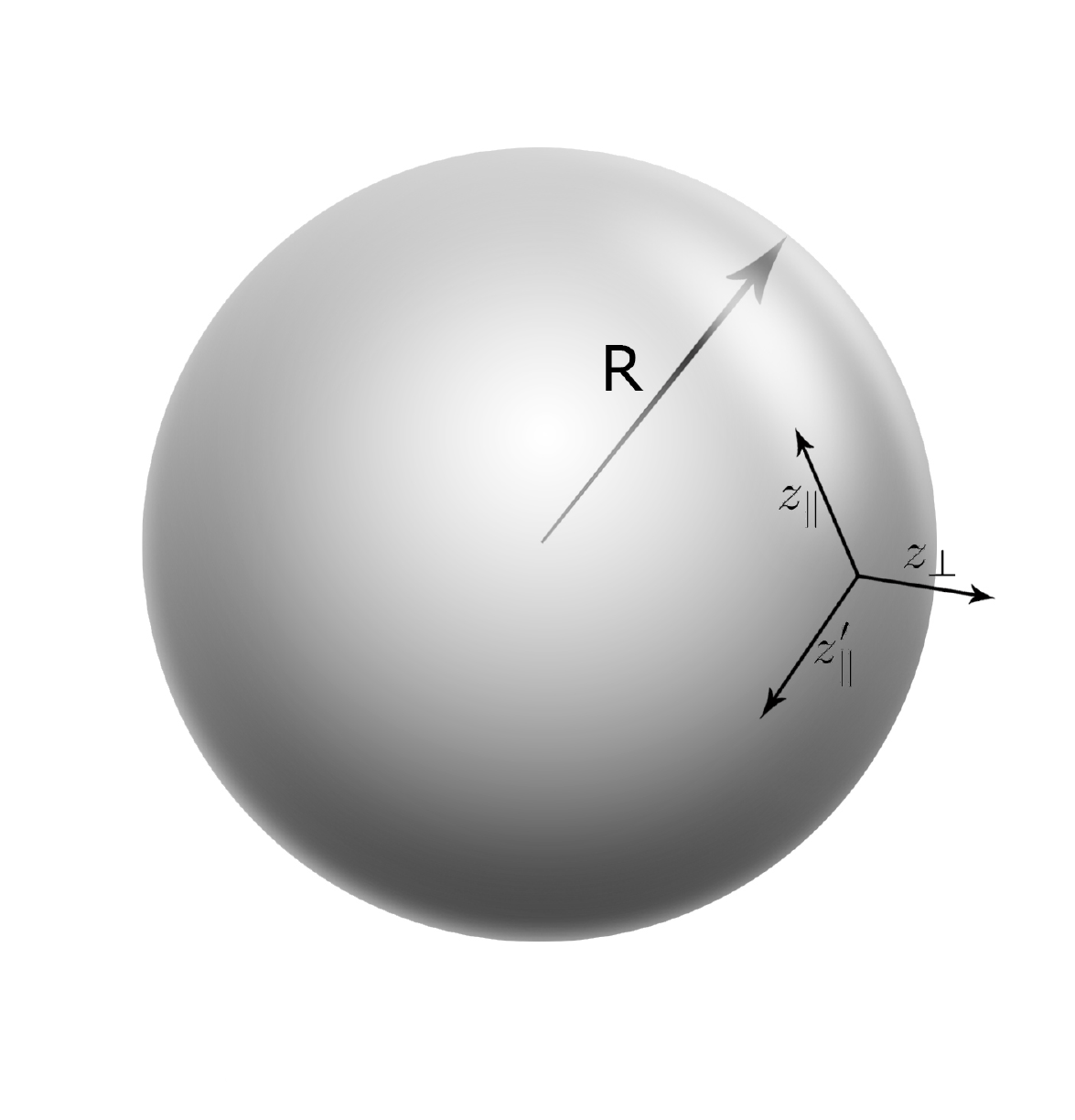}}
	\caption{Coordinate system in planar-wall approximation.}
	\label{planar-wall_coordinates}
\end{figure}

As was mentioned in introduction here we will restrict ourselves with the so called planar-wall approximation, which is good one in the case of large bubble radius $R$, see Fig. \ref{planar-wall_coordinates}. Here, the coordinates $\bf z_{\|}$ are parallel to the bubble surface, while  $z_{\bot}$ is the one orthogonal to it.  

To find a Green function in a bounce background we need to solve the following inhomogeneous Klein-Gordon equation
\begin{equation}
(\triangle^{(4)} + U''(\varphi)) G(\varphi ; x, x') = \delta^{(4)} (x - x')\, .
\end{equation}
To do that, it is convenient to perform Fourier transform with respect to coordinates parallel to the bubble
\begin{eqnarray} 
G(\varphi ; x, x')= \int\frac{d^{3}\mathbf{k}}{(2\pi)^{3}}e^{i\mathbf{k}(\mathbf{z_{\|}}-\mathbf{z_{\|}'})}G(\varphi ; z_{\bot},z_{\bot}',\mathbf{k})
\end{eqnarray}
Then the  Green function $G(\varphi ; z_{\bot},z_{\bot}',\mathbf{k})$ satisfies the equation  ($z, z' = z_{\bot},z'_{\bot}$): 
\begin{eqnarray} 
\label{gen_eq_green_z}
\left[-\frac{d^2}{dz^2}+\mathbf{k}^2+U''(\varphi)\right]G(\varphi ; z,z',\mathbf{k})=\delta(z-z')
\end{eqnarray}
Making change of variables $x=\tanh(\gamma z)$, $y=\tanh(\gamma z')$ we have $\varphi=vx$ ($v^2=12\gamma^2/\lambda$),
\begin{eqnarray} 
U''(\varphi)=-\mu^2+\frac{\lambda}{2}\varphi^2=4\gamma^2-6\gamma^2(1-x^2)
\end{eqnarray}
and the above equation takes the form
\begin{eqnarray} 
\label{gen_eq_green_x}
\left[\frac{d}{dx}(1-x^2)\frac{d}{dx}-\frac{m^2}{1-x^2}+6\right]G(x,y,\mathbf{k})=-\frac{\delta(x-y)}{\gamma}
\end{eqnarray}
where $m=\frac{1}{\gamma}\sqrt{4\gamma^2+\mathbf{k}^2}$. Here and below to shorten notation we will write $G(x,y,\mathbf{k})$ instead of $G(\varphi ; x,y,\mathbf{k})$. The equation \eqref{gen_eq_green_x} is solved differently depending on whether $m>2$ or $m=2$. When  $m>2$ the solution of homogeneous equation is given by 
\begin{eqnarray} 
G(x,y,\bk)=C_1(y)P_2^m(x)+C_2(y)Q_2^m(x)
\end{eqnarray}
where $P_2^m(x)$ and $Q_2^m(x)$ are the associated Legendre functions of the first and second kind, respectively. Applying the boundary conditions 
\begin{enumerate}
	\item[1)] $G(x,y,\bk)\rightarrow 0$ when $ x,y\rightarrow \pm 1 $
	\item[2)] continuity at $x=y$ 
	\item[3)] jump of the derivative at $x=y$: $\left.\frac{\partial}{\partial{x}}G^{x>y}(x,y,\bk)\right|_{y=x}-\left.\frac{\partial}{\partial{x}}G^{y<x}(x,y,\bk)\right|_{y=x}=-\frac{1}{\gamma(1-x^2)}$
\end{enumerate}
we get
\begin{eqnarray} 
G(x,y,\bk)=\theta(x-y)\frac{\pi}{2\gamma\sin m\pi}P^{-m}_{2}(x)P^{m}_{2}(y)+(x\leftrightarrow y)
\end{eqnarray}
Finally, employing the representation of associated Legendre function of the first kind in terms of Jacobi polynomials
\begin{equation}
P_n^m (z) = \left(\frac{z+1}{z-1}\right)^{\frac{m}{2}}(n-m+1)_m P_n^{(-m, m)}(z)
\end{equation} 
and using the fact that for $n=2$ the polynomial expansion of the latter terminates
\begin{equation}
P_2^{(\pm m, \mp m)} (z) = \frac{1}{2} \left[(1\pm m)(2\pm m) - 3 (2\pm m)(1-z) + 3 (1-z)^2\right]
\end{equation}
we get \cite{GreenFunctionScalarQFT}:
\begin{multline}
\label{GeneralGF}
G (x, y, \bk) = \, \frac{1}{2\gamma m} \Bigg\{ \theta (x-y) \left(
\frac{1-x}{1+x}
\right)^{\frac{m}{2}} \left(
\frac{1+y}{1-y}
\right)^{\frac{m}{2}} \left(
1 - 3 \frac{(1-x)(1+m+x)}{(1+m)(2+m)}
\right) \nonumber \\
 \times \left(
1 - 3 \frac{(1-y)(1-m+y)}{(1-m)(2-m)}
\right) + (x\to y)
\Bigg\}\, . 
\end{multline}
In the case $m=2$ the differential operator acting on Green function in \eqref{gen_eq_green_x} has zero mode and its inversion is consistently defined only on the subspace of functions orthogonal to this zero mode (Fredholm alternative). The corresponding equation for $m=2$ was already considered in \cite{Olejnik,MaximGreenFunctionm2} and is obtained by adding the product of properly normalized zero modes\footnote{The normalization factor $\sqrt{\frac{3\gamma}{4}}$ could be obtained for example from the condition $G(1,1,0)=G_{FV}(1,1,0)$.} ($\varphi_0=\sqrt{\frac{3\gamma}{4}}\frac{1}{\cosh^2(\gamma z)}$) to the right-hand side of the equation \eqref{gen_eq_green_z}, so that
\begin{eqnarray} 
\label{gen_eq_green_z_k0}
\left[-\frac{d^2}{dz^2}+U''(\varphi)\right]G(z,z',0)=\delta(z-z')-\frac{3\gamma}{4\cosh^2(\gamma z)\cosh^2(\gamma z')}
\end{eqnarray}
and equation \eqref{gen_eq_green_x} becomes
\begin{eqnarray} 
\label{gen_eq_green_x_k0}
\left[\frac{d}{dx}(1-x^2)\frac{d}{dx}-\frac{2^2}{1-x^2}+6\right]G(x,y,0)=\frac{-\delta(x-y)}{\gamma}+\frac{3}{4\gamma}(1-y^2)
\end{eqnarray}
The solution of this equation at $x\ne y$ is given by
\begin{multline} 
G(x,y,0)=\frac{1}{8\gamma}(1-y^2)\left(1+\frac{1}{1-x^2}\right)+C_1(y)(1-x^2)+
\\
+C_2(y)\left(\frac{3}{4}(1-x^2)\log\frac{1+x}{1-x}+\frac{x}{1-x^2}+\frac{3}{2}x\right)
\end{multline}
or in other form
\begin{eqnarray} 
G(x,y,0)=\frac{1}{8\gamma}(1-y^2)\left(1+\frac{1}{1-x^2}\right)+C_1(y)P_2^2(x)+C_2(y)Q_2^2(x)
\end{eqnarray}
In this case $G(x,y,k)$ must satisfy the following boundary conditions
\begin{enumerate}
	\item[1)] $G(x,y,k)\rightarrow 0$ when $ x,y\rightarrow \pm 1 $
	\item[2)] continuity at $x=y$
	\item[3)] orthogonality of $G(x,y,0)$ to the zero mode
	\begin{equation} 
	\int\limits_{-\infty}^{\infty}\frac{G(z,z',0)}{\cosh^2(\gamma z)}dz\sim \int\limits_{-1}^1G(x,y,0)dx=0
	\end{equation}
\end{enumerate}
which allow us to fix functions $C_{1,2}(y)$ and the result reads 
\begin{multline}
\label{Green0} 
G(x,y,0)=\frac{g(x,y)}{4\gamma}\left\{2-xy+\frac{|x-y|}{4}(11-3xy)+(x-y)^2\right\} \\ +\frac{3}{32\gamma}(1-x^2)(1-y^2)\left(\log g(x,y)-\frac{11}{3}\right)
\end{multline}
with
\begin{eqnarray}
\label{Green0_free}  
g(x,y)=\frac{1-|x-y|-xy}{1+|x-y|-xy}
\end{eqnarray}
Similarly in the case of false vacuum we have\footnote{Note that the differential operator acting on Green function in false-vacuum  does not contain zero modes, and therefore there is no need to consider the special case for $m = 2$. }
\begin{equation}
G_{FV} (x, y, \bk) = \frac{1}{2\gamma m} \Bigg\{ \theta (x-y) \left(
\frac{1-x}{1+x}
\right)^{\frac{m}{2}} \left(
\frac{1+y}{1-y}
\right)^{\frac{m}{2}}  + (x\to y)
\Bigg\}
\end{equation}
Transforming this expression back to $z$, $z'$ variables we get
\begin{equation} 
\label{ZGreen}
G_{FV}(z,z',\mathbf{k})=\frac{e^{-m(\bk)|z-z'|}}{2 m(\bk)}\, ,\quad m(\bk) = \sqrt{\hat{m}^2+\bk^2}\, , \quad \hat{m}^2 = -\mu^2 + v^2/2 = 4\gamma^2
\end{equation}
Now, if we take the Fourier transforms in variables $z$, $z'$  we recover ordinary  momentum space propagator we used to
\begin{equation}
G_{FV}(k) = \frac{1}{k^2 + \hat{m}^2}\, .
\end{equation}
Here, momentum $k$ is four-dimensional already. This later property is of great importance to us as it allows us to use for computations in false vacuum ordinary four-dimensional propagators.

\subsection{One-loop expression}\label{OneLoopDecay}

To get one-loop expression for false vacuum decay rate we need to evaluate difference of two traces \eqref{funcdet1loop}. In this work to regulate ultraviolet divergences we will use the dimensional regularization. Using the latter together with heat kernel method \cite{heatkernel-usermanual} we get\footnote{See \cite{GreenFunctionScalarQFT} for a similar derivation in a cut-off regularization}  
\begin{equation} 
I_{\unren}^{(1)}=\frac{1}{2}\int\limits_0^{\infty}\frac{d\tau}{\tau}\int\frac{d^{(d-1)}\mathbf{k}}{(2\pi)^{d-1}}\int d^{(d-1)}\mathbf{z_{\|}}\int d z_{\bot}\mathcal{L}^{-1}_s[G(z_{\bot},z_{\bot},\bk_s)- \\
G_{FV}(z_{\bot},z_{\bot},\bk_s)](\tau)
\end{equation}
Here $\bk_s$ denotes the substitution $\bk^2 \to \bk^2 + s$, $d=4-2\ep$ and  $\mathcal{L}^{-1}_s$ is the inverse Laplace transform with respect to $s$. The inverse Laplace transform and all integrals except over $\tau$ variable could be easily taken and the expression for $I^{(1)}$ takes the form 
\begin{eqnarray} 
I_{\unren}^{(1)}=\frac{1}{4\gamma^4}S_b\lambda(4\pi)^{-\frac{d-1}{2}}\int\limits_0^{\infty}d\tau\left[\frac{\gamma}{2}e^{-3\tau\gamma^2}\tau^{5/2+ \ep}\left(\erf(\sqrt{\tau}\gamma)+e^{3\tau\gamma^2}\erf(2\sqrt{\tau}\gamma)\right)\right]
\end{eqnarray}
To perform this final integration it is convenient to use the series representation of the error function
\begin{eqnarray} 
\erf(a)=\frac{2}{\sqrt{\pi}}e^{-a^2}\sum\limits_{l=0}^{\infty}\frac{2^l a^{2l+1}}{(2l+1)!!}
\end{eqnarray}
and after a term-by-term integration of the resulting series we get
\begin{eqnarray} 
I_{\unren}^{(1)}=\frac{1}{4\sqrt{\pi}}S_b\lambda(4\pi)^{-\frac{d-1}{2}}\gamma^{-2\ep}\sum\limits_{l=0}^{\infty}\frac{2^{2-2\ep-l}(1+2^{2l+1})\Gamma(l-1+\ep)}{(2l+1)!!}
\end{eqnarray}
The only functions with $\ep$ singularities are $\Gamma(\ep-1)$ and $\Gamma(\ep)$, so that the pole is generated only by first two terms in the series above. The remaining part of this series can be easily summed by putting $\ep=0$, which is allowed because this part is not multiplied by the functions containing singularities.
\begin{eqnarray} 
\sum\limits_{l=2}^{\infty}\frac{2^{2-l}(1+2^{2l+1})\Gamma(l-1)}{(2l+1)!!}=\frac{2\pi}{\sqrt{3}}
\end{eqnarray}
Finally, gathering everything together for the unrenormalized one-loop contribution we get
\begin{eqnarray} 
I_{\unren}^{(1)}=-S_b\left(\frac{3\lambda(e^{-\gamma_E} \pi \gamma^{-2})^{\ep}}{16\pi^2}\right)\left(\frac{1}{\ep}+2-\frac{\pi}{3\sqrt{3}}\right)
\end{eqnarray}
To get finite expression we need to account for ultraviolet renormalization, which is usual procedure in renormalizable quantum field theories. The needed counterterms could be conveniently introduced with the following additional interaction terms in the lagrangian
\begin{equation}
\LL_{\text{counterterms}} = \frac{1}{2}\delta\mu^2\Phi^2 + \frac{\delta\lambda}{4!}\Phi^4 + \frac{\delta Z}{2}(\partial_{\mu}\Phi)^2
\end{equation} 
The mass and coupling counterterms could be determined by considering renormalization of effective Coleman-Weinberg (CW) potential, while wave function renormalization is determined through renormalization of two-point function. In $\MSbar$ scheme at one-loop order we get ($\gE$ is the Euler constant)
\begin{align}
\delta\mu^{2 (1)} = & \, -\frac{\lambda\mu^2}{2(4\pi)^2\ep}(4 \pi e^{-\gamma_E})^{\ep}\, , \\
\delta\lambda^{(1)} = & \, \frac{3\lambda^2}{2(4\pi)^2\ep}(4 \pi e^{-\gamma_E})^{\ep}\, , \\
\delta Z^{(1)} = &\, 0\, ,
\end{align} 
which is in agreement with previously known results \cite{FunctionalMethodsPerturbationTheory,ScalingBehaviorPhi4,AnApproachEffectivePotential}. Then adding action counterterm
\begin{equation}
\delta S^{(1)} = \int d^4 x \left\{
\frac{1}{2}\delta\mu^2 (\phi^2 - v^2) + \frac{1}{4!}\delta\lambda (\phi^4 - v^4) + \frac{1}{2}\delta Z (\partial_{\mu}\phi)^2
\right\} = \frac{3S_b\lambda}{(4\pi)^2\ep}(4 \pi e^{-\gamma_E})^{\ep}
\end{equation}
the final expression for $I^{(1)}$ takes the form   
\begin{equation}
I^{(1)} = \frac{3 S_b\lambda}{(4\pi)^2}\left[\frac{\pi}{3\sqrt{3}}-2+\log\left(\frac{4\gamma^2}{\mu_{\MSbar}^2}\right)\right]\, .
\end{equation}
where $\mu_{\MSbar}$ is $\MSbar$ renormalization scale.  If we would use the Coleman-Weinberg renormalization scheme prescription, than at one-loop we would recover the result of \cite{GreenFunctionScalarQFT}. Note, that in \cite{GreenFunctionScalarQFT} the authors used regularization by cut-off. However, the final result in this particular scheme is independent of regularization prescription.  

\subsection{Two-loop corrections}\label{TwoLoopDecay}

The evaluation of higher order corrections for false vacuum decay is similar to quantum mechanical case considered in \cite{VacuumDecayQM}, see also  \cite{Instantons2loop,Olejnik,Instantons3loop,Instantons3loop-SineGordon,QuantumThermalFluctuationsQM} for calculation of other quantities. There are however technical complications related to the dimension of spacetime our quantum field theory lives in and the need for renormalization. The required Feynman rules needed for the computation of partition function around false vacuum and bounce solution are given by
\begin{align}
\vcenter{\hbox{\includegraphics[width=0.08\textwidth]{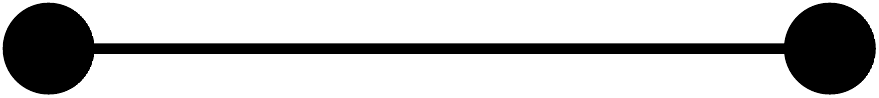}}}=\int\frac{d^{(d-1)}\mathbf{k}}{(2\pi)^{d-1}}e^{i\mathbf{k}(\mathbf{z_{\|}}-\mathbf{z_{\|}'})}G(z_{\bot},z_{\bot}',\mathbf{k})
\nonumber
\end{align}
\begin{align}
\vcenter{\hbox{\includegraphics[width=0.08\textwidth]{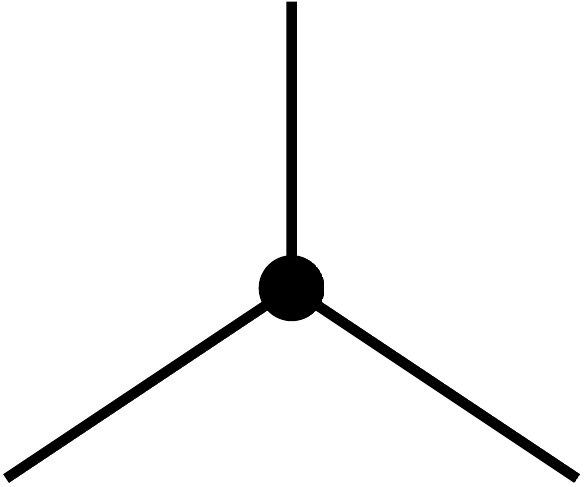}}}=2\gamma\sqrt{3\lambda}x\, ,\qquad &  \vcenter{\hbox{\includegraphics[width=0.08\textwidth]{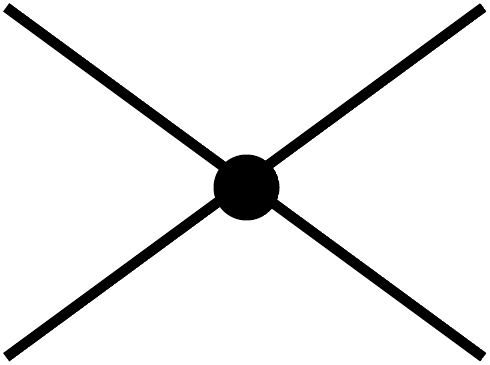}}}=-\lambda, \qquad &\vcenter{\hbox{\includegraphics[width=0.08\textwidth]{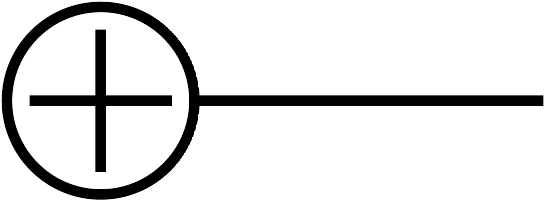}}}=\frac{2x(1-x^2)}{S_b^2} \nonumber 
\end{align}
\begin{align}
\vcenter{\hbox{\includegraphics[width=0.14\textwidth]{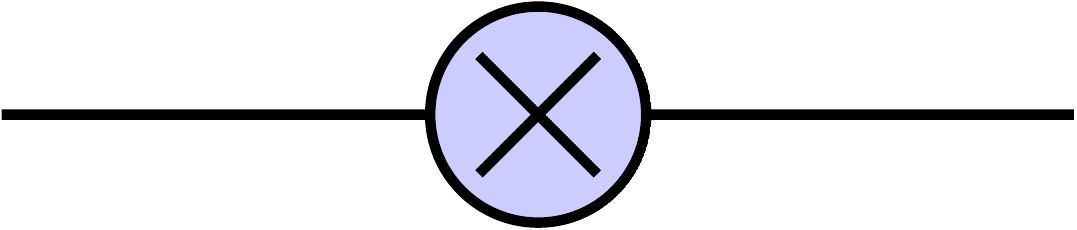}}}=-\delta\mu^2-\frac{\delta\lambda}{2}v^2x^2+2\gamma^2(3x^2-1)\delta Z\,  
\end{align}
\begin{align}
\vcenter{\hbox{\includegraphics[width=0.08\textwidth]{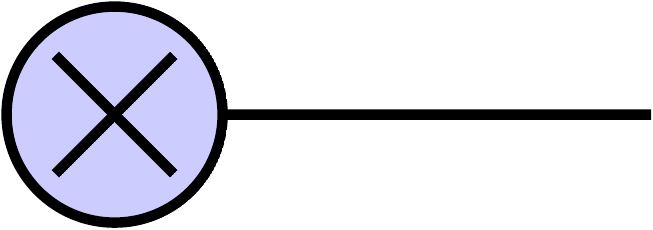}}}=vx\left(\delta\mu^2+\frac{\delta\lambda}{3!}v^2x^2-2\gamma^2\delta Z(2x^2-1)\right), \nonumber 
\end{align}
where $x=\tanh(\gamma z)$.

\begin{figure}[h]
	\center{\includegraphics[width=0.8\textwidth]{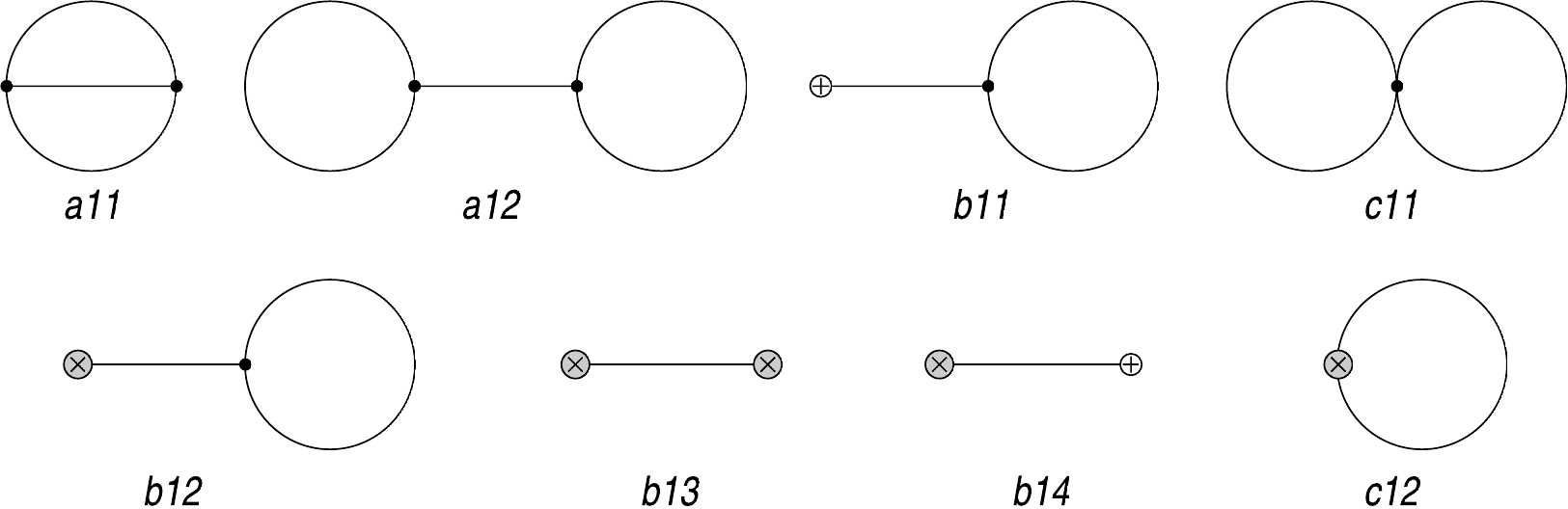}}
	\caption{Two loop Feynman diagrams.}
	\label{FD}
\end{figure}

The vertex with a plus sign inside a circle is a tadpole vertex coming from integration measure\footnote{See \cite{VacuumDecayQM} for its derivation in the case of quantum mechanics.}, while all other vertexes come from lagrangian. At two-loop order we need to calculate diagrams presented in Fig. \ref{FD}

All diagrams except sunset diagram $a11$ could be calculated more or less straightforwardly.  For example for diagram $a12$ we have 
\begin{multline} 
I_{a12}=\frac{2\gamma\sqrt{3\lambda}}{8}\int d z_{\bot}\int d z'_{\bot}\int d^{(d-1)}\mathbf{z_{\|}}\int d^{(d-1)}\mathbf{z'_{\|}}\int\frac{d^{d-1}\mathbf{k_1}}{(2\pi)^{d-1}}\int\frac{d^{d-1}\mathbf{k_2}}{(2\pi)^{d-1}}\int\frac{d^{d-1}\mathbf{k_3}}{(2\pi)^{d-1}}\times
\nonumber \\
\times\left\{G(z_{\bot},z_{\bot},\mathbf{k_1})G(z_{\bot},z'_{\bot},\mathbf{k_2})G(z'_{\bot},z'_{\bot},\mathbf{k_3})\tanh(\gamma z_{\bot})\tanh(\gamma z'_{\bot})\right.-\nonumber \\
\left.-G_{FV}(z_{\bot},z_{\bot},\mathbf{k_1})G_{FV}(z_{\bot},z'_{\bot},\mathbf{k_2})G_{FV}(z'_{\bot},z'_{\bot},\mathbf{k_3}) \right\}e^{i\mathbf{k_2 (z_{\|}-z'_{\|})}}
\end{multline}
Going from $z_{\bot}$, $z'_{\bot}$ to $x=\tanh (\gamma z_{\bot})$, $y=\tanh (\gamma z'_{\bot})$ variables  and taking all integrations except over $x$ and $y$ the expression for $I_{a12}$ takes the form
\begin{multline} 
I_{a12}=S_c^d\frac{3\lambda^2}{8\gamma^3}\int\limits_{-1}^1\int\limits_{-1}^1\frac{dxdy}{(1-x^2)(1-y^2)}\left\{xyG(x)G(y)G(x,y,0)-G^2(1)G_{FV}(x,y,0)\right\}
\end{multline}
Finally, using the integral expressions
\begin{multline}
G (x) = \int\frac{d^{d-1}\bk}{(2\pi)^{d-1}} G(z_{\bot}, z_{\bot}, \bk) = \frac{\gamma^{2-2\ep} e^{-\ep\gE}}{8\pi^{2-\ep}}\Bigg\{ \frac{1-3 x^2}{\ep} + 4 + x^2 (-6 + \sqrt{3}\pi (x^2-1)) \\
+ \ep \Big[ 10 - 12 x^2 + \frac{\pi^2}{12}(1-3 x^2) - \frac{\sqrt{3}\pi}{2} x^2 (x^2-1) (\log 3 - 4) \\
+ 3 i\sqrt{3} x^2 (x^2 - 1)\Big\{ \Li2 \Big(\frac{3-i\sqrt{3}}{6}\Big) - \Li2 \Big(\frac{3+i\sqrt{3}}{6}\Big)\Big\}\Big]\Bigg\}
\end{multline}
and
\begin{equation}
G_{FV} (x) = \int\frac{d^{d-1}\bk}{(2\pi)^{d-1}} G_{FV}(z_{\bot}, z_{\bot}, \bk) = G (1) = -\frac{\gamma^{2-2\ep}e^{-\ep\gE}}{4\pi^{2-\ep}}\left\{
\frac{1}{\ep} + 1 + \ep \left(1+\frac{\pi^2}{12}\right)
\right\}
\end{equation}
we get
\begin{multline} 
I_{a12}=\frac{S_c\lambda^2(e^{-\gamma_E} \pi \gamma^{-2})^{2\ep}}{2048\pi^4}\left[-\frac{39}{\ep^2}-\frac{96+\frac{58\pi}{5\sqrt{3}}}{\ep}-180-\frac{81\pi^2}{14}+\frac{44\sqrt{3}\pi}{5}-\right. 
\nonumber \\
-\left.\frac{29\sqrt{3}}{15}\left\{\pi\log 3-6i \Li2 \left(\frac{3-i\sqrt{3}}{6}\right)+6i \Li2 \left(\frac{3+i\sqrt{3}}{6}\right)\right\}\right]
\end{multline}
This example shows the use of the expression for Green function  $G(x,y,0)$ in the  special case with $m=2$, see section \ref{GreenFunctionsubsection}\footnote{In \cite{GreenFunctionScalarQFT} this problem was solved in another way by dividing the Green function for general $m$ (\ref{GeneralGF}) into  odd and even parts. Then, it was shown that the odd part containing infinity cancels out in actual calculations.  Both these methods give the same result (we checked it explicitly). Nevertheless we believe that our method is more mathematically correct as in this case Green function does not contain any unnatural divergences from the very beginning.}. The values of other diagrams could be found in a mathematica file accompanying the article and \ref{sunset-appendix} contains details of sunset diagram evaluation.

The two-loop counterterms are again determined from the renormalization of the effective potential and two-point function. In $\MSbar$ scheme at two-loop order we get
\begin{align}
\delta\mu^{2 (2)} = & \, -\frac{\lambda^2\mu^2(2-\ep)}{4(4\pi)^4\ep^2}(4 \pi e^{-\gamma_E})^{2\ep}\, , \\
\delta\lambda^{(2)} = & \, \frac{3\lambda^3(3-2\ep)}{4(4\pi)^4\ep^2}(4 \pi e^{-\gamma_E})^{2\ep} \, , \\
\delta Z^{(2)} = &\, -\frac{\lambda^2(4 \pi e^{-\gamma_E})^{2\ep}}{24(4\pi)^4\ep}\, ,
\end{align}    
which are again in agreement with previously obtained results \cite{FunctionalMethodsPerturbationTheory,ScalingBehaviorPhi4,AnApproachEffectivePotential}. Summing contribution of all diagrams\footnote{Diagrams with tadpole vertex coming from integration measure do not contribute in a planar thin wall approximation considered in this paper.} and adding two-loop action counterterm
\begin{equation} 
\delta S^{(2)}=-\frac{S_b\lambda^2(72-53\ep)}{4(4\pi)^4\ep^2}(4 \pi e^{-\gamma_E})^{2\ep}
\end{equation}
we finally get\footnote{Note, that the equation (\ref{decay_final1}) should be considered as the preferred definition for the false vacuum decay rate, see  \cite{Instantons2loop,Instantons3loop,Instantons3loop-SineGordon,VacuumDecayQM} for more details. On the other hand, the form (\ref{decay_final2}) is more convenient for carrying out the renormalization procedure.} 
\begin{eqnarray}
\label{decay_final1}  
\frac{\Gamma}{V}=\left(\frac{S_b}{2\pi\hbar}\right)^2\frac{(2\gamma)^5R}{\sqrt{3}}\exp(-\frac{1}{\hbar}S_b+ I^{(1)})\left\{1+\hbar I^{(2)}+\mathcal{O}(\hbar^2)\right\}
\end{eqnarray}
or
\begin{eqnarray}
\label{decay_final2}  
\frac{\Gamma}{V}=\left(\frac{S_b}{2\pi\hbar}\right)^2\frac{(2\gamma)^5R}{\sqrt{3}}\exp\big(-\frac{1}{\hbar}S_b + I^{(1)} + \hbar I^{(2)} +\mathcal{O}(\hbar^2)\big)\, , 
\end{eqnarray}
where 
\begin{multline}
I^{(2)} = \frac{S_b\lambda^2}{8\pi^4}\left[-\frac{3}{4}+\frac{7\sqrt{3}\pi}{160} - \frac{197\pi^2}{8960}-\frac{142-3\sqrt{3}\pi}{384}\log\left(\frac{4\gamma^2}{\mu_{\MSbar}^2}\right)+\frac{9}{256}\log^2\left(\frac{4\gamma^2}{\mu_{\MSbar}^2}\right)\right.
\nonumber \\
-\left.\frac{3\sqrt{3}}{320}\left\{\pi\log 3-6i \Li2 \left(\frac{3-i\sqrt{3}}{6}\right)+6i \Li2 \left(\frac{3+i\sqrt{3}}{6}\right)\right\}+s_0\right]
\end{multline}
Here $s_0$ is the finite part of the sunset diagram $a11$, see \ref{sunset-appendix}. We would like
to note that part of two-loop corrections related to bounce renormalization  considered in \cite{GreenFunctionScalarQFT} is contained in the sum of our diagrams $a12$, $b12$ and $b13$. These diagrams are obtained by inserting tadpole contribution for scalar field self-energy found in \cite{GreenFunctionScalarQFT} into corresponding one-loop diagram for false vacuum decay rate. Numerically, this contribution is of the order of 10\% of full two-loop result.

\begin{figure}[h]
	\center{\includegraphics[width=0.5\textwidth]{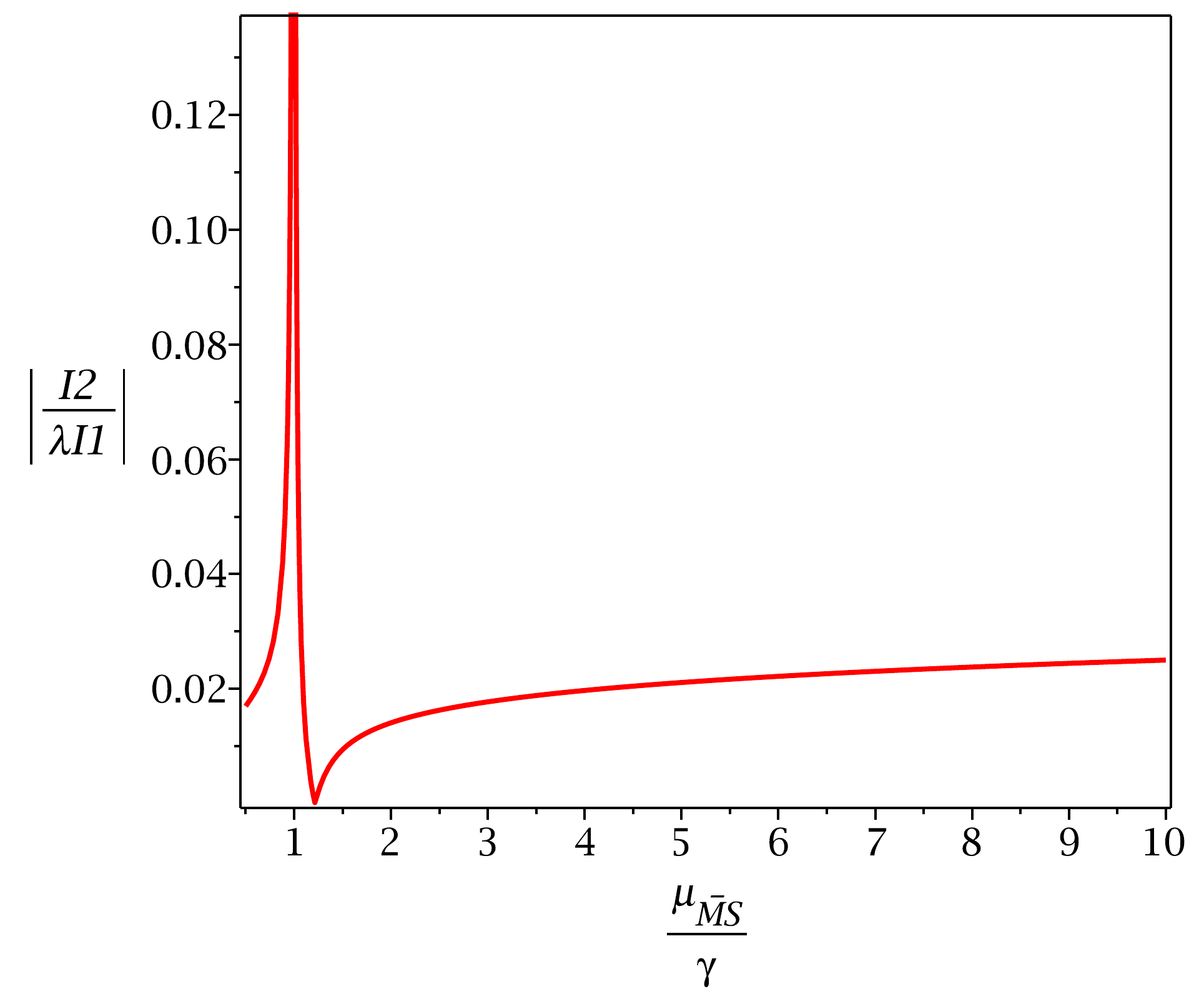}}
	\caption{The ratio of $I^{(2)}$ to $\lambda I^{(1)}$ as a function of $\mu_{\MSbar}/\gamma$.}
	\label{ratio}
\end{figure} 

Vacuum decay rate is a measurable quantity and as such it should not depend from the scale choice. This means that our result need to satisfy the following equation:
\begin{equation}
\frac{d}{d \log \mu_{\MSbar}}\left(-\frac{1}{\hbar} S_b + I^{(1)} + \mathcal{O}(\hbar)\right)=0
\end{equation}
We have verified that it is indeed the case up to order $\hbar$ by expressing $S_b$ and $I^{(1)}$ in terms of the Lagrangian parameters $\gamma,~g,~\lambda$ and applying to them the renormalization group equations \cite{SMtunneling4}:
\begin{equation}
\frac{d\gamma}{d \log \mu_{\MSbar}}=\hbar\ \frac{\gamma\lambda}{32\pi^2}
\end{equation}
\begin{equation}
\frac{dg}{d \log\mu_{\MSbar}}=\hbar\ \frac{3g\lambda}{16\pi^2}
\end{equation}
\begin{equation}
\frac{d\lambda}{d \log\mu_{\MSbar}}=\hbar\ \frac{3\lambda^2}{16\pi^2}
\end{equation}
Note, that since we are using thing wall approximation ($R\to\infty$, $g\to 0$) we should leave only leading order terms in coupling $g$ in renormalization group equations. Moreover, it turns out that within thin-wall approximation the knowledge of one-loop beta-functions is sufficient to check the coefficients in front of logarithms with $\mu_{\MSbar}$ dependence at two-loop order also. So, we see that there is no renormalization-scale uncertainty in this method\footnote{As usual there could be some residual scale dependence left at one perturbation order higher.}.

Finally, in order to understand the qualitative significance of our result, let us consider the ratio of $I^{(2)}$ to $\lambda I^{(1)}$ as a function of $\mu_{\MSbar}/\gamma$. From Fig. \ref{ratio} we see that at reasonable values of $\mu_{\MSbar} \gtrsim 2\gamma$, where neither $I^{(1)}$ or $I^{(2)}$ are close to zero, this ratio is about $0.02-0.03$. This means that the two loop correction have very little impact on the vacuum decay rate and can be safely neglected whenever we do not interested in accuracy greater than $2-3\%$.

\section{Conclusion}\label{Conclusion}

In this work we computed for the first time two-loop quantum corrections to false vacuum decay in a four-dimensional scalar field theory with cubic and quartic potential. Using planar thin wall approximation we were able to get analytical expression for the latter. The results obtained employ dimensional regularization and $\MSbar$ renormalization scheme. It is shown that the obtained decay rate is independent from renormalization scale variation. It turns out that two-loop corrections is approximately $2-3\%$ of one-loop result. So, we may conclude that one-loop approximation accounting for prefactor to exponent of the bounce action is a well defined approximation to false vacuum decay in the model considered. In a subsequent paper we are planning to present the generalization of the obtained results to Coleman-Weinberg (CW) scheme and cut-off regularization.

\section*{Acknowledgements}

The authors would like to thank O.~Veretin and A.~Bednyakov for interesting and stimulating discussions. This work was supported by RFBR grants \# 17-02-00872, \# 16-02-00943 and contract \# 02.A03.21.0003 from 27.08.2013 with Russian Ministry of Science and Education. 

\appendix

\section{Sunset diagram evaluation}
\label{sunset-appendix}

The expression for sunset diagram $a11$ in Fig. \ref{FD} is given by 
\begin{multline} 
I_{sun}=\frac{(2\gamma\sqrt{3\lambda})^2}{12}\int d z_{\bot}\int d z'_{\bot}\int d^{(d-1)}\mathbf{z_{\|}}\int d^{(d-1)}\mathbf{z'_{\|}}\int\frac{d^{d-1}\mathbf{k_1}}{(2\pi)^{d-1}}\int\frac{d^{d-1}\mathbf{k_2}}{(2\pi)^{d-1}}\int\frac{d^{d-1}\mathbf{k_3}}{(2\pi)^{d-1}}\times
\nonumber \\
\times\Big\{G(z_{\bot},z'_{\bot},\mathbf{k_1})G(z_{\bot},z'_{\bot},\mathbf{k_2})G(z_{\bot},z'_{\bot},\mathbf{k_3})\tanh(\gamma z_{\bot})\tanh(\gamma z'_{\bot}) \nonumber \\
- G_{FV}(z_{\bot},z'_{\bot},\mathbf{k_1})G_{FV}(z_{\bot},z'_{\bot},\mathbf{k_2})G_{FV}(z_{\bot},z'_{\bot},\mathbf{k_3}) \Big\}e^{i\mathbf{(k_1+k_2+k_3) (z_{\|}-z'_{\|})}}
\end{multline}
Performing change of variables $x = \tanh (\gamma z_{\bot})$, $y = \tanh (\gamma z'_{\bot})$ and taking integrals over coordinates parallel to the bubble $\mathbf{z_{\|}}$ and $\mathbf{z'_{\|}}$ we get 
\begin{multline}
I_{sun} = -\frac{\lambda^2}{2} S_b \gamma^{-4\ep} \int_{-1}^1\int_{-1}^1 \frac{dx dy}{(1-x^2)(1-y^2)}\int\frac{d^{d-1}\bk_1}{(2\pi)^{d-1}}\int\frac{d^{d-1}\bk_2}{(2\pi)^{d-1}} \\ \times
\Big\{ 
x y G(x, y, \bk_1) G(x, y, \bk_2) G(x, y, \bk_3) - G_{FV}(x, y, \bk_1) G_{FV}(x, y, \bk_2) G_{FV}(x, y, \bk_3) 
\Big\} \\
\end{multline}
with $\bk_3 = \bk_1 + \bk_2$. To evaluate this integral it is convenient to modify Green function $G(x,y,\bk)$ as 
\begin{multline}
G (x, y, \bk) = \, \frac{1}{2\gamma m} \Bigg\{ \theta (x-y) \left(
\frac{1-x}{1+x}
\right)^{\frac{m}{2}} \left(
\frac{1+y}{1-y}
\right)^{\frac{m}{2}} \left(
1 - 3 \frac{(1-x)(1+m+x)}{(\frac{\gamma_m}{\gamma_M}+m)(\frac{2\gamma_m}{\gamma_M}+m)}
\right) \nonumber \\
\times \left(
1 - 3 \frac{(1-y)(1-m+y)}{(\frac{\gamma_m}{\gamma_M}-m)(\frac{2\gamma_m}{\gamma_M}-m)}
\right) + (x\to y)
\Bigg\}\, , 
\end{multline}
where $m = \frac{1}{\gamma_M}\sqrt{\bk^2 + 4\gamma_M^2}$. The original integral is recovered at $\gamma_m = \gamma_m = \gamma$. However for $\gamma_m < \gamma_M$ we may evaluate the integral as a series in $\frac{\gamma_m}{\gamma_M}$ and provided it is convergent evaluate its value at $\frac{\gamma_m}{\gamma_M} = 1$. It turns out that it is indeed the case. Moreover to obtained desired expansion we may use a strategy of regions, see \cite{StrategyRegions1,StrategyRegions2,StrategyRegions3,StrategyRegions4} and references therein. In our particular case only one region contributes, namely the one with $\bk_1 \sim \bk_2\sim \gamma_M$ and we may just perform the usual Taylor expansion of the integrand in $\gamma_m$. Then, the integrand contains the factors 
\begin{align}
\left(\frac{1-x}{1+x}\right)^{M/2} \left(\frac{1+y}{1-y}\right)^{M/2}\quad
\text{or} \quad \left(\frac{1-y}{1+y}\right)^{M/2} \left(\frac{1+x}{1-x}\right)^{M/2}\, ,
\end{align}
where $M = m_1 + m_2 + m_2$ and $m_i = \frac{1}{\gamma_M}\sqrt{\bk_i^2 + 4\gamma_M^2}$. In our region for large $\gamma_M$ $M\to \infty$ and we may use saddle point approximation to evaluate integrals over $x$, $y$ variables.  That is we set
\begin{align}
y = x + \frac{z}{M}
\end{align}
and Taylor expand integrands at $z=0$. Now, taking into account that ($M\to\infty$):
\begin{equation}
\begin{cases}
z \in (-\infty , 0) & x > y\\
z \in (0, \infty) & y > x
\end{cases}
\end{equation}
and taking integrals over $z$ we will for example obtain
\begin{align}
&\int_{-1}^1 \frac{d y}{1-y^2} \theta (x-y) \left(\frac{1-x}{1+x}\right)^{M/2} \left(\frac{1+y}{1-y}\right)^{M/2} = \frac{1}{M}\, , \\
&\int_{-1}^1 \frac{d y}{1-y^2} \theta (y-x) \left(\frac{1-y}{1+y}\right)^{M/2} \left(\frac{1+x}{1-x}\right)^{M/2} = \frac{1}{M} \, .
\end{align}
Finally taking integrals over $x$ and $y$ we get
\begin{multline}
I_{sun} = \frac{\lambda^2}{4}S_b \gamma^{-4\ep} \frac{1}{(2\pi)^{2 d-2}}\int d^{d-1}\bk_1 \int d^{d-1}\bk_2 \frac{1}{M m_1 m_2 m_3} \\
\times \left\{
1 - \frac{1}{m_1^2} - \frac{1}{m_2^2} - \frac{1}{m_3^2} - \frac{1}{m_1 M} -\frac{1}{m_2 M} - \frac{1}{m_3 M} + \frac{2}{3 M^2} + \ldots 
\right\} \label{sunset-series}
\end{multline}
Here $\dots$ denote higher order terms in the expansion. To further evaluate integrals over $\bk_1$ and $\bk_2$ we derived Mellin-Barnes representation for the integral
\begin{equation}
\int d^d k_1 \int d^d k_2 \frac{1}{m_1^{a_1} m_2^{a_2} m_3^{a_3} M^{a_4}}\, ,
\end{equation}
where $d=3-2\ep$, $m_i = (k_i^2+4)^{1/2}$, $M=m_1+m_2+m_3$ as before and $a_i$ are arbitrary indexes. The latter is given by 
\begin{multline}
\int d^d k_1 \int d^d k_2 \frac{1}{m_1^{a_1} m_2^{a_2} m_3^{a_3} M^{a_4}} = \frac{\pi^d 2^{2 d-a_1-a_2-a_3-a_4}}{(2\pi i)^3\Gamma (a_4)\Gamma\left(\frac{d}{2}\right)}\int_{-i\infty}^{i\infty} dz_1 \int_{-i\infty}^{i\infty} dz_2 \int_{-i\infty}^{i\infty} dz_3  \\
\times \frac{\Gamma (-z_1) \Gamma (-z_2) \Gamma (-z_3) \Gamma (a_4 + z_1+z_2)\Gamma\left(\frac{a_3+a_4+z_1+z_2+2z_3}{2}\right)}{\Gamma \left(
\frac{a_3+a_4+z_1+z_2}{2}	
\right)\Gamma\left(\frac{a_1-z_1}{2}\right)\Gamma\left(\frac{a_2-z_2}{2}\right)} \\
\times \frac{\Gamma \left(\frac{a_1+a_3+a_4+z_2+2 z_3-d}{2}\right)\Gamma \left(\frac{a_2+a_3+a_4+z_1+2 z_3-d}{2}\right)\Gamma\left(
\frac{d-a_3-a_4-z_1-z_2-2z_3}{2}	
\right)\Gamma\left(\frac{a_1+a_2+a_3+a_4+2z_3-2d}{2}\right)}{\Gamma\left(
\frac{a_1+a_2+z_1+z_2}{2} + a_3 + a_4 + 2z_3 -d
\right)}\, .
\end{multline}
The resulting Mellin-Barnes integrals where evaluated numerically with the help of \cite{Czakon}. Finally for the sunset diagram we got 
\begin{equation} 
\label{sunset_final}
I_{sun}=\frac{S_c\lambda^2(e^{-\gamma_E} \pi \gamma^{-2})^{2\ep}}{8\pi^4}\left[\frac{9}{64\ep^2}+\frac{s_{-1}}{\ep}+s_0\right]\, ,
\end{equation}
where $s_{-1} = 0.39522$ and $s_0\approx 0.71$. The value of $s_{-1} = 0.39522$ which we got from the first $20$ terms of the series\footnote{We have written explicitly only two of them in \eqref{sunset-series}} \eqref{sunset-series} is actually several percent less then the exact value which could be found from the cancellation of $1/\ep$ poles in the process of renormalization 
\begin{equation}
s_{-1} = \frac{197}{384} - \frac{3\sqrt{3}\pi}{160}\approx 0.410995
\end{equation}
which is explained by slow convergences of the mentioned series. It could be certainly further improved but this goes beyond the goal of the present paper. 

\section*{References}

\bibliography{litr}

\begin{thebibliography}{10}

\bibitem{SMtunneling1}
G.~Isidori, G.~Ridolfi, and A.~Strumia, ``{On the metastability of the standard
  model vacuum},'' {\em Nucl. Phys.}, vol.~B609, pp.~387--409, 2001.

\bibitem{SMtunneling2}
Z.~Lalak, M.~Lewicki, and P.~Olszewski, ``{Higher-order scalar interactions and
  SM vacuum stability},'' {\em JHEP}, vol.~05, p.~119, 2014.

\bibitem{SMtunneling3}
A.~D. Plascencia and C.~Tamarit, ``{Convexity, gauge-dependence and tunneling
  rates},'' {\em JHEP}, vol.~10, p.~099, 2016.

\bibitem{SMtunneling4}
M.~Endo, T.~Moroi, M.~M. Nojiri, and Y.~Shoji, ``{Renormalization-Scale
  Uncertainty in the Decay Rate of False Vacuum},'' {\em JHEP}, vol.~01,
  p.~031, 2016.

\bibitem{SMtunneling5}
Z.~Lalak, M.~Lewicki, and P.~Olszewski, ``{Gauge fixing and renormalization
  scale independence of tunneling rate in Abelian Higgs model and in the
  standard model},'' {\em Phys. Rev.}, vol.~D94, no.~8, p.~085028, 2016.

\bibitem{SMtunneling6}
O.~Czerwińska, Z.~Lalak, M.~Lewicki, and P.~Olszewski, ``{The impact of
  non-minimally coupled gravity on vacuum stability},'' {\em JHEP}, vol.~10,
  p.~004, 2016.

\bibitem{SMtunneling7}
J.~R. Espinosa, M.~Garny, T.~Konstandin, and A.~Riotto, ``{Gauge-Independent
  Scales Related to the Standard Model Vacuum Instability},'' {\em Phys. Rev.},
  vol.~D95, no.~5, p.~056004, 2017.

\bibitem{SMtunneling8}
M.~Endo, T.~Moroi, M.~M. Nojiri, and Y.~Shoji, ``{On the Gauge Invariance of
  the Decay Rate of False Vacuum},'' {\em Phys. Lett.}, vol.~B771,
  pp.~281--287, 2017.

\bibitem{SMtunneling9}
M.~Endo, T.~Moroi, M.~M. Nojiri, and Y.~Shoji, ``{False Vacuum Decay in Gauge
  Theory},'' {\em JHEP}, vol.~11, p.~074, 2017.

\bibitem{SMtunneling10}
A.~Andreassen, W.~Frost, and M.~D. Schwartz, ``{Scale Invariant Instantons and
  the Complete Lifetime of the Standard Model},'' {\em Phys. Rev.}, vol.~D97,
  no.~5, p.~056006, 2018.

\bibitem{SMtunneling11}
S.~Chigusa, T.~Moroi, and Y.~Shoji, ``{State-of-the-Art Calculation of the
  Decay Rate of Electroweak Vacuum in the Standard Model},'' {\em Phys. Rev.
  Lett.}, vol.~119, no.~21, p.~211801, 2017.

\bibitem{SMtunneling12}
S.~Chigusa, T.~Moroi, and Y.~Shoji, ``{Decay Rate of Electroweak Vacuum in the
  Standard Model and Beyond},'' 2018.

\bibitem{SMtunneling13}
B.-H. Lee and W.~Lee, ``{Vacuum bubbles in a de Sitter background and black
  hole pair creation},'' {\em Class. Quant. Grav.}, vol.~26, p.~225002, 2009.

\bibitem{SMtunneling14}
B.-H. Lee, C.~H. Lee, W.~Lee, and C.~Oh, ``{Instanton solutions mediating
  tunneling between the degenerate vacua in curved space},'' {\em Phys. Rev.},
  vol.~D82, p.~024019, 2010.

\bibitem{SMmetastability1}
J.~Elias-Miro, J.~R. Espinosa, G.~F. Giudice, G.~Isidori, A.~Riotto, and
  A.~Strumia, ``{Higgs mass implications on the stability of the electroweak
  vacuum},'' {\em Phys. Lett.}, vol.~B709, pp.~222--228, 2012.

\bibitem{SMmetastability2}
G.~Degrassi, S.~Di~Vita, J.~Elias-Miro, J.~R. Espinosa, G.~F. Giudice,
  G.~Isidori, and A.~Strumia, ``{Higgs mass and vacuum stability in the
  Standard Model at NNLO},'' {\em JHEP}, vol.~08, p.~098, 2012.

\bibitem{SMmetastability3}
F.~Bezrukov, M.~{\relax Yu}. Kalmykov, B.~A. Kniehl, and M.~Shaposhnikov,
  ``{Higgs Boson Mass and New Physics},'' {\em JHEP}, vol.~10, p.~140, 2012.
\newblock [,275(2012)].

\bibitem{SMmetastability4}
S.~Alekhin, A.~Djouadi, and S.~Moch, ``{The top quark and Higgs boson masses
  and the stability of the electroweak vacuum},'' {\em Phys. Lett.}, vol.~B716,
  pp.~214--219, 2012.

\bibitem{SMmetastability5}
I.~Masina, ``{Higgs boson and top quark masses as tests of electroweak vacuum
  stability},'' {\em Phys. Rev.}, vol.~D87, no.~5, p.~053001, 2013.

\bibitem{SMmetastability6}
D.~Buttazzo, G.~Degrassi, P.~P. Giardino, G.~F. Giudice, F.~Sala, A.~Salvio,
  and A.~Strumia, ``{Investigating the near-criticality of the Higgs boson},''
  {\em JHEP}, vol.~12, p.~089, 2013.

\bibitem{SMmetastability7}
J.~R. Espinosa, G.~F. Giudice, E.~Morgante, A.~Riotto, L.~Senatore, A.~Strumia,
  and N.~Tetradis, ``{The cosmological Higgstory of the vacuum instability},''
  {\em JHEP}, vol.~09, p.~174, 2015.

\bibitem{SMmetastability8}
A.~V. Bednyakov, B.~A. Kniehl, A.~F. Pikelner, and O.~L. Veretin, ``{Stability
  of the Electroweak Vacuum: Gauge Independence and Advanced Precision},'' {\em
  Phys. Rev. Lett.}, vol.~115, no.~20, p.~201802, 2015.

\bibitem{Coleman1}
S.~R. Coleman, ``{The Fate of the False Vacuum. 1. Semiclassical Theory},''
  {\em Phys. Rev.}, vol.~D15, pp.~2929--2936, 1977.
\newblock [Erratum: Phys. Rev.D16,1248(1977)].

\bibitem{Coleman2}
C.~G. Callan, Jr. and S.~R. Coleman, ``{The Fate of the False Vacuum. 2. First
  Quantum Corrections},'' {\em Phys. Rev.}, vol.~D16, pp.~1762--1768, 1977.

\bibitem{Coleman3}
S.~R. Coleman and F.~De~Luccia, ``{Gravitational Effects on and of Vacuum
  Decay},'' {\em Phys. Rev.}, vol.~D21, p.~3305, 1980.

\bibitem{Kobzarev}
I.~{\relax Yu}. Kobzarev, L.~B. Okun, and M.~B. Voloshin, ``{Bubbles in
  Metastable Vacuum},'' {\em Sov. J. Nucl. Phys.}, vol.~20, pp.~644--646, 1975.
\newblock [Yad. Fiz.20,1229(1974)].

\bibitem{KleinertPathIntegrals}
H.~Kleinert, ``{Path Integrals in Quantum Mechanics, Statistics, Polymer
  Physics, and Financial Markets},'' 2004.

\bibitem{IntroQuantumMechanics}
H.~J.~W. Müller-Kirsten, {\em {Introduction to Quantum Mechanics}}.
\newblock World Scientific, 2012.

\bibitem{QFTCriticalPhenomena}
J.~Zinn-Justin, ``{Quantum field theory and critical phenomena},'' {\em Int.
  Ser. Monogr. Phys.}, vol.~113, pp.~1--1054, 2002.

\bibitem{InstantonsLargeN}
M.~Mariño, {\em {Instantons and Large N}}.
\newblock Cambridge University Press, 2015.

\bibitem{WeinbergClassicalSolutions}
E.~J. Weinberg, {\em {Classical solutions in quantum field theory}}.
\newblock Cambridge Monographs on Mathematical Physics, Cambridge University
  Press, 2012.

\bibitem{PrecisionDecayRates}
A.~Andreassen, D.~Farhi, W.~Frost, and M.~D. Schwartz, ``{Precision decay rate
  calculations in quantum field theory},'' {\em Phys. Rev.}, vol.~D95, no.~8,
  p.~085011, 2017.

\bibitem{DirectApproachQuantumTunneling}
A.~Andreassen, D.~Farhi, W.~Frost, and M.~D. Schwartz, ``{Direct Approach to
  Quantum Tunneling},'' {\em Phys. Rev. Lett.}, vol.~117, no.~23, p.~231601,
  2016.

\bibitem{heatkernel-usermanual}
D.~V. Vassilevich, ``{Heat kernel expansion: User's manual},'' {\em Phys.
  Rept.}, vol.~388, pp.~279--360, 2003.

\bibitem{CalculationsExternalFields}
V.~A. Novikov, M.~A. Shifman, A.~I. Vainshtein, and V.~I. Zakharov,
  ``{Calculations in External Fields in Quantum Chromodynamics. Technical
  Review},'' {\em Fortsch. Phys.}, vol.~32, p.~585, 1984.

\bibitem{MassiveContributionsQCDtunneling}
O.-K. Kwon, C.-k. Lee, and H.~Min, ``{Massive field contributions to the QCD
  vacuum tunneling amplitude},'' {\em Phys. Rev.}, vol.~D62, p.~114022, 2000.

\bibitem{DunneFunctionalDeterminants}
G.~V. Dunne, ``{Functional determinants in quantum field theory},'' {\em J.
  Phys.}, vol.~A41, p.~304006, 2008.

\bibitem{KleinertChervyakov1}
H.~Kleinert and A.~Chervyakov, ``Functional determinants from wronski green
  functions,'' {\em Journal of Mathematical Physics}, vol.~40, no.~11,
  pp.~6044--6051, 1999.

\bibitem{KleinertChervyakov2}
H.~Kleinert and A.~Chervyakov, ``{Simple explicit formulas for Gaussian path
  integrals with time dependent frequencies},'' {\em Phys. Lett.}, vol.~A245,
  pp.~345--357, 1998.

\bibitem{GreenFunctionScalarQFT}
B.~Garbrecht and P.~Millington, ``{Green’s function method for handling
  radiative effects on false vacuum decay},'' {\em Phys. Rev.}, vol.~D91,
  p.~105021, 2015.

\bibitem{GarbrechtMillington2}
B.~Garbrecht and P.~Millington, ``{Self-consistent solitons for vacuum decay in
  radiatively generated potentials},'' {\em Phys. Rev.}, vol.~D92, p.~125022,
  2015.

\bibitem{GarbrechtMillington3}
B.~Garbrecht and P.~Millington, ``{Self-consistent radiative corrections to
  false vacuum decay},'' {\em J. Phys. Conf. Ser.}, vol.~873, no.~1, p.~012041,
  2017.

\bibitem{GarbrechtMillington4}
B.~Garbrecht and P.~Millington, ``{Fluctuations about the Fubini-Lipatov
  instanton for false vacuum decay in classically scale invariant models},''
  2018.

\bibitem{GenfaldYaglom}
I.~M. Gelfand and A.~M. Yaglom, ``{Integration in functional spaces and it
  applications in quantum physics},'' {\em J. Math. Phys.}, vol.~1, p.~48,
  1960.

\bibitem{KirstenMcKane1}
K.~Kirsten and A.~J. McKane, ``{Functional determinants by contour integration
  methods},'' {\em Annals Phys.}, vol.~308, pp.~502--527, 2003.

\bibitem{KirstenMcKane2}
K.~Kirsten and A.~J. McKane, ``{Functional determinants for general
  Sturm-Liouville problems},'' {\em J. Phys.}, vol.~A37, pp.~4649--4670, 2004.

\bibitem{ABCinstantons}
A.~I. Vainshtein, V.~I. Zakharov, V.~A. Novikov, and M.~A. Shifman, ``{ABC's of
  Instantons},'' {\em Sov. Phys. Usp.}, vol.~25, p.~195, 1982.
\newblock [,201(1981)].

\bibitem{InstantonsQCD}
T.~Schäfer and E.~V. Shuryak, ``{Instantons in QCD},'' {\em Rev. Mod. Phys.},
  vol.~70, pp.~323--426, 1998.

\bibitem{LecturesInstantons}
S.~Vandoren and P.~van Nieuwenhuizen, ``{Lectures on instantons},'' 2008.

\bibitem{Instantons2loop}
A.~A. Aleinikov and E.~V. Shuryak, ``{Instantons in quantum mechanics. Two loop
  effects},'' {\em Yad. Fiz.}, vol.~46, pp.~122--129, 1987.

\bibitem{Olejnik}
S.~Olejnik, ``{Do nongaussian effects decrease tunneling probabilities? Three
  loop instanton density for the double well potential},'' {\em Phys. Lett.},
  vol.~B221, pp.~372--376, 1989.

\bibitem{Instantons3loop}
M.~A. Escobar-Ruiz, E.~Shuryak, and A.~V. Turbiner, ``{Three-loop Correction to
  the Instanton Density. I. The Quartic Double Well Potential},'' {\em Phys.
  Rev.}, vol.~D92, no.~2, p.~025046, 2015.
\newblock [Erratum: Phys. Rev.D92,no.8,089902(2015)].

\bibitem{Instantons3loop-SineGordon}
M.~A. Escobar-Ruiz, E.~Shuryak, and A.~V. Turbiner, ``{Three-loop Correction to
  the Instanton Density. II. The Sine-Gordon potential},'' {\em Phys. Rev.},
  vol.~D92, no.~2, p.~025047, 2015.

\bibitem{QuantumThermalFluctuationsQM}
M.~A. Escobar-Ruiz, E.~Shuryak, and A.~V. Turbiner, ``{Quantum and thermal
  fluctuations in quantum mechanics and field theories from a new version of
  semiclassical theory},'' {\em Phys. Rev.}, vol.~D93, no.~10, p.~105039, 2016.

\bibitem{VacuumDecayQM}
M.~A. Bezuglov and A.~I. Onishchenko, ``{Radiative corrections to false vacuum
  decay in quantum mechanics},'' {\em Phys. Rev.}, vol.~D96, no.~3, p.~036001,
  2017.

\bibitem{EulerHeisenberg1}
G.~V. Dunne and C.~Schubert, ``{Two loop selfdual Euler-Heisenberg Lagrangians.
  1. Real part and helicity amplitudes},'' {\em JHEP}, vol.~08, p.~053, 2002.

\bibitem{EulerHeisenberg2}
G.~V. Dunne and C.~Schubert, ``{Two loop selfdual Euler-Heisenberg Lagrangians.
  2. Imaginary part and Borel analysis},'' {\em JHEP}, vol.~06, p.~042, 2002.

\bibitem{EulerHeisenberg3}
G.~V. Dunne, ``{Heisenberg-Euler effective Lagrangians: Basics and
  extensions},'' in {\em From fields to strings: Circumnavigating theoretical
  physics. Ian Kogan memorial collection (3 volume set)} (M.~Shifman,
  A.~Vainshtein, and J.~Wheater, eds.), pp.~445--522, 2004.

\bibitem{Tseytlin1}
R.~Roiban, A.~Tirziu, and A.~A. Tseytlin, ``{Two-loop world-sheet corrections
  in AdS(5) x S**5 superstring},'' {\em JHEP}, vol.~07, p.~056, 2007.

\bibitem{Tseytlin2}
R.~Roiban and A.~A. Tseytlin, ``{Spinning superstrings at two loops:
  Strong-coupling corrections to dimensions of large-twist SYM operators},''
  {\em Phys. Rev.}, vol.~D77, p.~066006, 2008.

\bibitem{NonperturbativeMethodsReview}
R.~F. Dashen, B.~Hasslacher, and A.~Neveu, ``{Nonperturbative Methods and
  Extended Hadron Models in Field Theory 1. Semiclassical Functional
  Methods},'' {\em Phys. Rev.}, vol.~D10, p.~4114, 1974.

\bibitem{MaximGreenFunctionm2}
M.~Bezuglov, ``{False vacuum decay in quantum mechanics and four dimensional
  scalar field theory},'' {\em EPJ Web Conf.}, vol.~177, p.~09001, 2018.

\bibitem{FunctionalMethodsPerturbationTheory}
J.~Iliopoulos, C.~Itzykson, and A.~Martin, ``{Functional Methods and
  Perturbation Theory},'' {\em Rev. Mod. Phys.}, vol.~47, p.~165, 1975.

\bibitem{ScalingBehaviorPhi4}
J.~C. Collins, ``{Scaling behavior of phi-to-the-4 theory and dimensional
  regularization},'' {\em Phys. Rev.}, vol.~D10, pp.~1213--1218, 1974.

\bibitem{AnApproachEffectivePotential}
R.~Grigjanis, R.~Kobes, and Y.~Fujimoto, ``{An approach to the effective
  potential},'' {\em Can. J. Phys.}, vol.~64, pp.~537--545, 1986.

\bibitem{StrategyRegions1}
V.~A. Smirnov, ``{`Strategy of regions': Expansions of Feynman diagrams both in
  Euclidean and pseudo-Euclidean regimes},'' in {\em {Proceedings, 5th
  International Symposium on Radiative Corrections - RADCOR 2000}}, 2001.

\bibitem{StrategyRegions2}
V.~A. Smirnov, ``{Problems of the strategy of regions},'' {\em Phys. Lett.},
  vol.~B465, pp.~226--234, 1999.

\bibitem{StrategyRegions3}
V.~A. Smirnov and E.~R. Rakhmetov, ``{The Strategy of regions for asymptotic
  expansion of two loop vertex Feynman diagrams},'' {\em Theor. Math. Phys.},
  vol.~120, pp.~870--875, 1999.
\newblock [Teor. Mat. Fiz.120,64(1999)].

\bibitem{StrategyRegions4}
M.~Beneke and V.~A. Smirnov, ``{Asymptotic expansion of Feynman integrals near
  threshold},'' {\em Nucl. Phys.}, vol.~B522, pp.~321--344, 1998.

\bibitem{Czakon}
M.~Czakon, ``{Automatized analytic continuation of Mellin-Barnes integrals},''
  {\em Comput. Phys. Commun.}, vol.~175, pp.~559--571, 2006.

\end{thebibliography}

\end{document}